\documentclass[preprintnumbers,amsmath,amssymbm,prd]{revtex4}
\usepackage{epsfig}
\usepackage{graphicx}
\usepackage{xcolor}

\begin{document}
\title{The third law of black hole thermodynamics and the mass-to-charge ratio of scalar fields}
\author{Shahar Hod$^1$ and Tsvi Piran$^2$}
\affiliation{$^1$The Ruppin Academic Center, Emeq Hefer 40250, Israel}
\affiliation{$^2$Racah Institute of Physics, The Hebrew University, Jerusalem 91904, Israel}
\date{\today}

\begin{abstract}
Kehle and Unger have recently shown that self-gravitating charged classical scalar fields can collapse to form 
extremal Reissner-Nordstr\"om (eRN) black holes in finite time, thus violating the third law of thermodynamics (TLT). 
Reall has subsequently proved that the formation of an eRN black hole 
in gravitational collapse is impossible if the mass-to-charge ratio of the field exceeds unity, $m/e\geq1$. 
In the present compact paper we conjecture, using quantum considerations, that the Reall bound is {\it not} sharp. 
In particular, we point out that the electric field of a charged system with charge $Q$ will spontaneously 
polarize the vacuum in the regime $(eQ/\hbar)^2>(mQ/\hbar)^2+1/4$, thereby creating pairs of charged particles that 
discharge the black hole and ensure the quantum validity of the TLT in this regime. 
Motivated by the TLT, we conjecture that the classical 
Reall bound can be strengthened such that classical considerations would prevent eRN black holes from forming dynamically in the regime $m/e\geq\sqrt{1-\big({{\hbar}/{2eQ}}\big)^2}$, 
where quantum polarization effects are unable to protect the TLT.
\end{abstract}
\bigskip
\maketitle

\section{Introduction}

The third law of thermodynamics (TLT) places a fundamental restriction on the attainment of absolute zero. 
In its standard formulation, no physically realizable process can bring a thermodynamic system to the zero-temperature 
state $T=0$ after a finite sequence of operations, or, equivalently, within a finite amount of time \cite{TL}. 
Whether analogous principle constrains the formation of zero-temperature (extremal) black holes is a particularly interesting question. 

The thermodynamic character of black holes emerged from the interplay between classical general relativity 
and quantum theory. Bekenstein argued in the 1970s that black holes should be assigned 
a nonzero entropy in a semiclassical description, 
with the entropy proportional to the horizon area and inversely proportional to the fundamental Planck area $\hbar$ 
(we adopt units in which $G=c=1$) \cite{Bk73,Bk74}. 
The work of Bardeen, Carter, and Hawking established a set of four mechanical relations 
for stationary black holes that closely parallel the ordinary laws of thermodynamics \cite{BCH}. 
The thermodynamic interpretation became compelling, however, only after Hawking's discovery that 
black holes emit thermal radiation with a quantum ($\hbar$-dependent) 
temperature which is proportional to their surface gravity \cite{H75}. 

The quantum nature of black-hole thermodynamics is particularly evident in the 
generalized second law (GSL). The classical area theorem, established in a series 
of works \cite{Ch70,CR71,Ha71}, states that the area of an event horizon 
cannot decrease under appropriate physical assumptions. Bekenstein \cite{Bk81,Bk83} 
and Unruh and Wald \cite{UW82,UW83} 
subsequently demonstrated that the corresponding generalized entropy principle requires quantum effects. 
Thus, although the classical geometry provides the basis for the area theorem, quantum physics is essential for the formulation and consistency of the GSL in situations involving black holes \cite{Bk81,Bk83,UW82,UW83}. 

A similar question arises for the third law. Building on classical ideas, Israel \cite{Is86} established a celebrated classical formulation of the black-hole third law, according to which an initially nonextremal black hole cannot be driven to extremality by a physically reasonable process in finite advanced time. 
The status of this result was recently reconsidered by Kehle and Unger \cite{KU} (see also \cite{Schn}). 
They constructed solutions of the classical Einstein-Maxwell-charged-massless-scalar system in which an initially non-extremal configuration evolves into an exactly extremal Reissner-Nordstr\"om black hole after a finite amount of advanced time. Since extremality corresponds to vanishing Hawking temperature, their construction appears to provide a counterexample to the black-hole version of the TLT. 
The apparent violation of the TLT identified by Kehle and Unger was subsequently extended to the coupled Einstein-Maxwell-charged-{\it massive}-scalar system in \cite{Gad,Lee}.

These developments raise a fundamental issue concerning the relation between classical gravitational dynamics and quantum black-hole thermodynamics. In particular, one may  ask whether the extremal configurations produced in the Kehle-Unger construction remain physically attainable once quantum effects associated with the charged matter fields are taken into account. In this work, we investigate this question precisely: 
Can quantum physics prevent the formation of extremal charged black holes suggested by the Kehle-Unger construction, thereby restoring the validity of the third law of black-hole thermodynamics?

\section{Quantum physics and the TLT}

Like the GSL, the TLT is fundamentally a {\it quantum} law \cite{Hodbo,HodQ2,Gru,Pes,HodQ3}. 
Quantum-mechanical effects should therefore play a crucial role in protecting its validity. 
In collapse scenarios involving self-gravitating charged fields, a potentially important protective mechanism 
is Schwinger-type pair production \cite{Schw1,Schw2,HodPir1}. This intrinsically quantum phenomenon can discharge the collapsing configuration and thereby prevent the formation of zero-temperature black holes.

In particular, as shown in \cite{Kt}, near-extremal Reissner-Nordstr\"om black holes are subject to spontaneous production of oppositely charged particles whenever
\begin{equation}\label{Eq1}
\bigg({{eQ}\over{\hbar}}\bigg)^2>\bigg({{mQ}\over{\hbar}}\bigg)^2+{1\over4}\  ,
\end{equation}
where $e$ and $m$ are, respectively, the electric charge and the proper mass of the field, 
and $Q$ is the electric charge of the black hole (we assume $Q>0$ without loss of generality). 
When this condition is met, the electric field of the black hole is sufficiently strong to induce vacuum pair production. The oppositely charged particles are driven in opposite directions by the electric field, with one member carrying 
charge away from the black hole. 
This process therefore provides a quantum mechanism for the discharge of the black hole. 
Note, in particular, that at the threshold (\ref{Eq1}), the absorbed quantum carries energy $-m$ and  
a negative electric charge $|q|>m$, thereby driving the charged system away from extremality.

Thus, in the dimensionless regime (\ref{Eq1}), the classical formation of extremal charged black holes 
through the collapse of charged massive scalar fields does not imply a violation of the TLT within the broader 
framework of a semi-classical theory of gravity.

\section{A conjectured bound on the mass-to-charge ratio of scalar fields in classical TLT-violating collapse scenarios}

While the quantum discharge phenomenon is effective in the regime (\ref{Eq1}), one may naturally ask whether 
the TLT can be violated outside this regime. 
Remarkably, classical effects appear to provide the necessary protection. 
Reall \cite{Reall1} has recently proved that, for sufficiently regular solutions, eRN black holes cannot 
form dynamically in finite time through gravitational collapse if
\begin{equation}\label{Eq2}
{{m}\over{e}}\geq1\qquad\Longrightarrow\qquad\text{No dynamically formed classical extremal black holes exist}. \
\end{equation}
It is presently unknown whether this bound is sharp. 

The arguments presented above lead us to conjecture that the Reall bound (\ref{Eq2}) is {\it not} sharp. 
More specifically, we conjecture that, in the Einstein-Maxwell-massive-scalar field theory, an 
extremal (zero-temperature) Reissner-Nordstr\"om black hole of charge $Q$ 
cannot form dynamically from a charged massive scalar field whose parameters satisfy 
\begin{equation}\label{Eq3}
\bigg({{eQ}\over{\hbar}}\bigg)^2\leq\bigg({{mQ}\over{\hbar}}\bigg)^2+{1\over4}
\qquad\Longrightarrow\qquad\text{No dynamically formed classical extremal black holes exist}. \
\end{equation}
Put differently, we conjecture that the classical Einstein-Maxwell-charged-massive-scalar field equations 
safeguard the TLT in the regime where quantum effects can no longer provide a mechanism 
for protecting this fundamental principle. 
Note that the bound (\ref{Eq3}) implies, in particular, that extremal 
Reissner-Nordstr\"om black holes cannot form dynamically from the collapse of massless scalar fields 
in the regime $eQ/\hbar\leq 1/2$ \cite{HodPir1}.

The conjectured bound (\ref{Eq3}) yields the $(eQ/\hbar)$-dependent mass-to-charge relation 
\begin{equation}\label{Eq4}
{{m}\over{e}}\geq\Big({{m}\over{e}}\Big)_{\text{max}}=\sqrt{1-\bigg({{\hbar}\over{2eQ}}\bigg)^2}
\qquad\Longrightarrow\qquad\text{No dynamically formed extremal black holes exist}, \
\end{equation}
which is in the spirit of, but slightly stronger than, the Reall bound (\ref{Eq2}). 
Incidentally, $\hbar$ appears in this classical formula because the electric charges $Q$ and $e$ have 
dimensions of length, so the introduction of $\hbar$ renders the physical quantity $eQ/\hbar$ dimensionless, 
making the connection between classical and quantum physics particularly transparent in the context of gravity. The $\hbar$ ``disappears", of course, when one adopts natural units with $G=c=\hbar=1$, as is done, 
for example, in \cite{KU,Reall1}. 

It is worthwhile to compare this conjectured limit with recent numerical results. 
Lee \cite{Lee} has recently demonstrated numerically the classical formation of extremal 
Reissner-Nordstr\"om black holes through the collapse of charged massive scalar fields 
with $m/e$ slightly below unity. 
In particular, Lee reported the following numerically computed maximum mass-to-charge ratios \cite{Lee}  
\begin{equation}\label{Eq5}
(m/e, eM/\hbar) = (0.9944, 46.80),\ (0.9961, 29.99),\ (0.9866, 59.99)\
\end{equation}
for the charged massive scalar fields that can produce extremal black holes 
using $C^1, C^2$, and $C^3$ gluing techniques, respectively.  

As the limiting values of $m/e$ are close to unity, one can expect at first glance that this provides a strong 
evidence that Reall's bound is sharp. 
However, while the numerical results presented by Lee \cite{Lee} respect the original 
Reall bound (\ref{Eq2}) they also respect our conjectured bound (\ref{Eq4}). 
In particular, using Eqs. (\ref{Eq4}) and (\ref{Eq5}) one finds the dimensionless ratios   
\begin{equation}\label{Eq6}
{{(m/e)_{\text{numerical}}}\over{(m/e)_{\text{max}}}} = 0.9945,\ 0.9962,\ 0.9866<1\
\end{equation}
between the numerically computed maximum mass-to-charge ratios of the scalar fields considered in \cite{Lee} 
and the TLT-motivated bound $({{m}/{e}})_{\text{max}}$ given by Eq. (\ref{Eq4}) [the ratios 
presented in Eq. (\ref{Eq6}) correspond, respectively, to charged massive scalar fields that dynamically 
form extremal black holes with $C^1, C^2$, and $C^3$ gluing]. 
It should be emphasized that a decisive test of the bound (\ref{Eq4}) 
requires collapse simulations with $eQ/\hbar=O(1)$ as done in \cite{Gad}. 

Although the difference between our conjectured classical bound (\ref{Eq4}) and the Reall bound (\ref{Eq2}) may be small for large black holes with $\hbar/eQ\ll1$, it has profound physical significance, as our conjectured bound preserves the TLT. 
We further conjecture that the bound (\ref{Eq4}) is sharp: that is,
extremal Reissner-Nordstr\"om black holes can only be formed classically from charged scalar fields with $m/e$ 
arbitrarily close to, but below, the threshold (\ref{Eq4}). 
If so, the classical theory would forbid the formation of extremal black holes precisely up to the point at which quantum pair production takes over, so that the TLT would be protected, without a gap, by classical dynamics on one side of the threshold and by quantum physics on the other.

\bigskip
\noindent
{\bf ACKNOWLEDGMENTS}
\bigskip

We would like to thank Don Page for helpful correspondence. 



\begin{thebibliography}{99}

\bibitem{TL} M. Bailyn, {\it A Survey of Thermodynamics}, 
American Institute of Physics, New York, ISBN 0-88318-797-3, page 342 (1994).

\bibitem{Bk73} J. D. Bekenstein, Phys. Rev. D {\bf 7}, 2333 (1973). 

\bibitem{Bk74} J. D. Bekenstein, Phys. Rev. D {\bf 9}, 3292 (1974). 

\bibitem{BCH} J. M. Bardeen, B. Carter, and S. W. Hawking, Comm. Math. Phys. {\bf 31}, 161 (1973). 

\bibitem{H75} S. W. Hawking, Comm. Math. Phys. {\bf 43}, 199 (1975).

\bibitem{Ch70} D. Christodoulou, Phys. Rev. Lett. {\bf 25}, 1596 (1970).

\bibitem{CR71} D.  Christodoulou and  R. Ruffini, Phys. Rev. D. {\bf  4}, 3552 (1971).

\bibitem{Ha71} S. W. Hawking, Phys. Rev. Lett. {\bf 26}, 1344 (1971). 

\bibitem{Bk81} J. D. Bekenstein, Phys. Rev. D {\bf 23}, 287 (1981).

\bibitem{Bk83} J. D. Bekenstein, Phys. Rev. D {\bf 27}, 2262 (1983).

\bibitem{UW82} W. G. Unruh and R. M. Wald, Phys. Rev. D {\bf 25}, 942 (1982).

\bibitem{UW83} W. G. Unruh and R. M. Wald, Phys. Rev. D {\bf 27}, 2271 (1983).

\bibitem{Is86} W. Israel, Phys. Rev. Lett. {\bf 57}, 397 (1986). 

\bibitem{KU} C. Kehle and R. Unger, arXiv:2211.15742.

\bibitem{Schn} B. Schneider, arXiv:2608.23355.

\bibitem{Gad} M. Gadioux, H. S. Reall, and J. E. Santos, arXiv:2512.10008.

\bibitem{Lee} J. Lee, arXiv:2609.22518.

\bibitem{Hodbo} S. Hod, Phys. Rev. D {\bf 75}, 064013 (2007) [arXiv:gr-qc/0611004]. 

\bibitem{HodQ2} S. Hod, Class. and Quant. Grav. {\bf 24}, 4235 (2007) [arXiv:0705.2306].

\bibitem{Gru} A. Gruzinov, arXiv:gr-qc/0705.1725.

\bibitem{Pes} A. Pesci, Class. Quantum Grav. {\bf 24}, 6219 (2007).

\bibitem{HodQ3} S. Hod, Phys. Lett. B {\bf 666}, 483 (2008) [arXiv:0810.5419]; 
S. Hod, Phys. Rev. D {\bf 78}, 084035 (2008) [arXiv:0811.3806]; 
S. Hod, Phys. Rev. D {\bf 80}, 064004 (2009) [arXiv:0909.0314]; 
S. Hod, Phys. Lett. B {\bf 823} , 136733 (2021) [arXiv:2202.01230].

\bibitem{Schw1} F. Sauter, Z. Phys. {\bf 69}, 742 (1931); W. Heisenberg and H. Euler,
Z. Phys. {\bf 98}, 714 (1936); J. Schwinger, Phys. Rev. {\bf 82}, 664 (1951).

\bibitem{Schw2} M. A. Markov and V. P. Frolov, Teor. Mat. Fiz. {\bf 3}, 3 (1970); 
W. T. Zaumen, Nature {\bf 247}, 531 (1974); 
B. Carter, Phys. Rev. Lett. {\bf 33}, 558 (1974); 
G. W. Gibbons, Comm. Math. Phys. {\bf 44}, 245 (1975); 
L. Parker and J. Tiomno, Astrophys. Journ. {\bf 178}, 809 (1972); 
T. Damour and R. Ruffini, Phys. Rev. Lett. {\bf 35}, 463 (1975).

\bibitem{HodPir1} S. Hod and T. Piran, arXiv:2609.16127.

\bibitem{Kt} C. M. Chen, S. P. Kim, I. C. Lin, J. R. Sun, and M. F. Wu, Phys. Rev. D {\bf 85}, 
124041 (2012) [arXiv:1202.3224]. 

\bibitem{Reall1} H. S. Reall, Phys. Rev. D {\bf 110}, 124059 (2024) [arXiv: 2410.11956]. 

\end{thebibliography}
\end{document}